\documentclass[11pt,a4paper]{article}
\usepackage{graphicx} % Required for inserting images

\usepackage[a4paper,margin=2cm]{geometry}
\usepackage{marginnote}
\usepackage{kvoptions}
\usepackage{multicol}

\usepackage{newtxtext}  % Times New Roman-like serif font

\usepackage{xcolor}
\usepackage{tcolorbox}
\definecolor{boxcolor}{RGB}{193, 229, 255}

\usepackage[T1]{fontenc}
\usepackage[utf8]{inputenc}

\usepackage{fancyhdr}
\usepackage{sectsty}
\allsectionsfont{\sffamily\bfseries}
\makeatletter
\renewcommand\section{%
  \@startsection{section}{1}{\z@}%
    {10pt \@plus 4pt \@minus 2pt}% beforeskip
    {1pt \@plus 1pt \@minus 1pt}% afterskip
    {\normalfont\large\bfseries\sffamily}%
}
\renewcommand\subsection{%
  \@startsection{subsection}{1}{\z@}%
    {10pt \@plus 4pt \@minus 2pt}% beforeskip
    { }% afterskip
    {\normalfont\large\bfseries\sffamily}%
}
\makeatother

\usepackage[numbers]{natbib}
\input{bib.def}

\usepackage{bibspacing}
\usepackage{bm}
\bibpunct[, ]{[}{]}{;}{n}{,}{,}
\renewcommand{\newblock}{}

\usepackage{graphicx}
\usepackage{caption}
\usepackage{url}
\usepackage{floatrow}
\usepackage{float}

\usepackage{setspace}
  {}           % end part

\usepackage{hanging}

\newcounter{affil}

\newcommand{\affilsup}[1]{\textsuperscript{#1}}

\newcommand{\useaffil}[1]{\affilsup{\ref{aff:#1}}}

\newcommand{\makeaffil}[2]{%
  \refstepcounter{affil}%
  \label{aff:#1}%
  \setstretch{0.7}
  \noindent\affilsup{\theaffil}\ {\small\itshape #2}\par%
}
\begin{document}
%----------------------------------------------------------------------
% --- TITLE PAGE --- 
%----------------------------------------------------------------------
\thispagestyle{empty} 
\begin{center} 
\textsf{Expanding Horizons}\\[1em]
    {\LARGE \textsf{\textbf{Millimeter-Wavelength Observations of the Active Sun:\\[0.5em] 
    \huge Unveiling the Origins of Space Weather}}}\\[2em]
\end{center}
\vspace{-3mm}
\normalsize 
    Sven Wedemeyer\useaffil{rocs}$^,$\useaffil{ita}, 
    Stefaan Poedts\useaffil{kuleuven}$^,$\useaffil{umcs},     
    Stanislav Gun\'ar\useaffil{cas},    
    Manuela Temmer\useaffil{graz}$^,$\useaffil{kanz}, 
    Astrid Veronig\useaffil{graz}$^,$\useaffil{kanz}, 
    Valery Nakariakov\useaffil{warw}, 
    Mats Kirkaune\useaffil{rocs}$^,$\useaffil{ita}, 
    Claudia Cicone\useaffil{ita}, 
    Stephen White\useaffil{airf}$^,$\useaffil{unm}, 
    Jasmina Magdalenić\useaffil{kuleuven}, 
    Roman Braj\v sa\useaffil{zag}, 
    Bart De Pontieu\useaffil{lmsal}$^,$\useaffil{rocs}$^,$\useaffil{ita}, 
    Maryam Saberi\useaffil{rocs}$^,$\useaffil{ita}, 
    Atul Mohan\useaffil{godd}$^,$\useaffil{caua}, 
    Davor Sudar\useaffil{zag}, 
    Galina Motorina \useaffil{pulk}$^,$\useaffil{iof}, 
    Maria Lukicheva \useaffil{mps}$^,$\useaffil{saor}, 
    Paulo Simões \useaffil{mack}$^,$\useaffil{glas}

% Affiliations 
\vspace{3mm}

\makeaffil{rocs}{Rosseland Centre for Solar Physics, University of Oslo, PO Box 1029, Blindern 0315 Oslo, Norway}

\makeaffil{ita}{Institute of Theoretical Astrophysics, University of Oslo, PO Box 1029, Blindern 0315 Oslo, Norway}

\makeaffil{kuleuven}{CmPA, Dept.\ of Mathematics, KU Leuven, Celestijnenlaan 200B, 3001 Leuven, Belgium} 

\makeaffil{umcs}{Institute of Physics, University of Maria Curie-Skłodowska, Pl.\ Marii Curie-Skłodowskiej 1, 20-031 Lublin, Poland} 

\makeaffil{cas}{Astronomical Institute, The Czech Academy of Sciences, 251 65 Ond\v rejov, Czech Republic} 

\makeaffil{graz}{Institute of Physics, University of Graz, Universitätsplatz 5, 8010 Graz, Austria} 

\makeaffil{kanz}{Kanzelh\"ohe Observatory for Solar and Environmental Research, University of Graz, Austria} 

\makeaffil{warw}{Centre for Fusion, Space and Astrophysics, Department of Physics, University of Warwick, Coventry CV4 7AL, UK} 

\makeaffil{airf}{Space Vehicles Directorate, Air Force Research Laboratory, Kirtland AFB, NM, USA}

\makeaffil{unm}{Department of Physics and Astronomy, University of New Mexico, Albuquerque, NM, 87106, USA}

\makeaffil{zag}{University of Zagreb, Faculty of Geodesy, Hvar Observatory, Ka\v ci\'ceva 26, 10000 Zagreb, Croatia} 

\makeaffil{lmsal}{Lockheed Martin Solar \& Astrophysics Laboratory, Palo Alto, CA 94304, USA}

\makeaffil{godd}{NASA Goddard Space Flight Center, Greenbelt, MD 20771, USA}

\makeaffil{caua}{The Catholic University of America, Washington, DC 20064, USA}

%Galina Motorina 
\makeaffil{pulk}{Central Astronomical Observatory at Pulkovo of Russian Academy of Sciences, St. Petersburg, 196140, Russia}

\makeaffil{iof}{Ioffe Institute, Polytekhnicheskaya, 26, St. Petersburg, 194021, Russia}

%Maria Lukicheva 
\makeaffil{mps}{Max Planck Institute for Solar System Research, Goettingen, Germany}

\makeaffil{saor}{Special Astrophysical Observatory, Nizhnii Arkhyz, Russia}

%Paulo Simões
\makeaffil{mack}{Center for Radio Astronomy and Astrophysics Mackenzie, School of Engineering, Mackenzie Presbyterian University, São Paulo, Brazil} 

\makeaffil{glas}{SUPA School of Physics and Astronomy, University of Glasgow, Glasgow G12 8QQ, UK}

% --- Nice front page figure --- 
\begin{figure}[b!]
\vspace{-5mm}
    \centering
    \includegraphics[width=17cm]{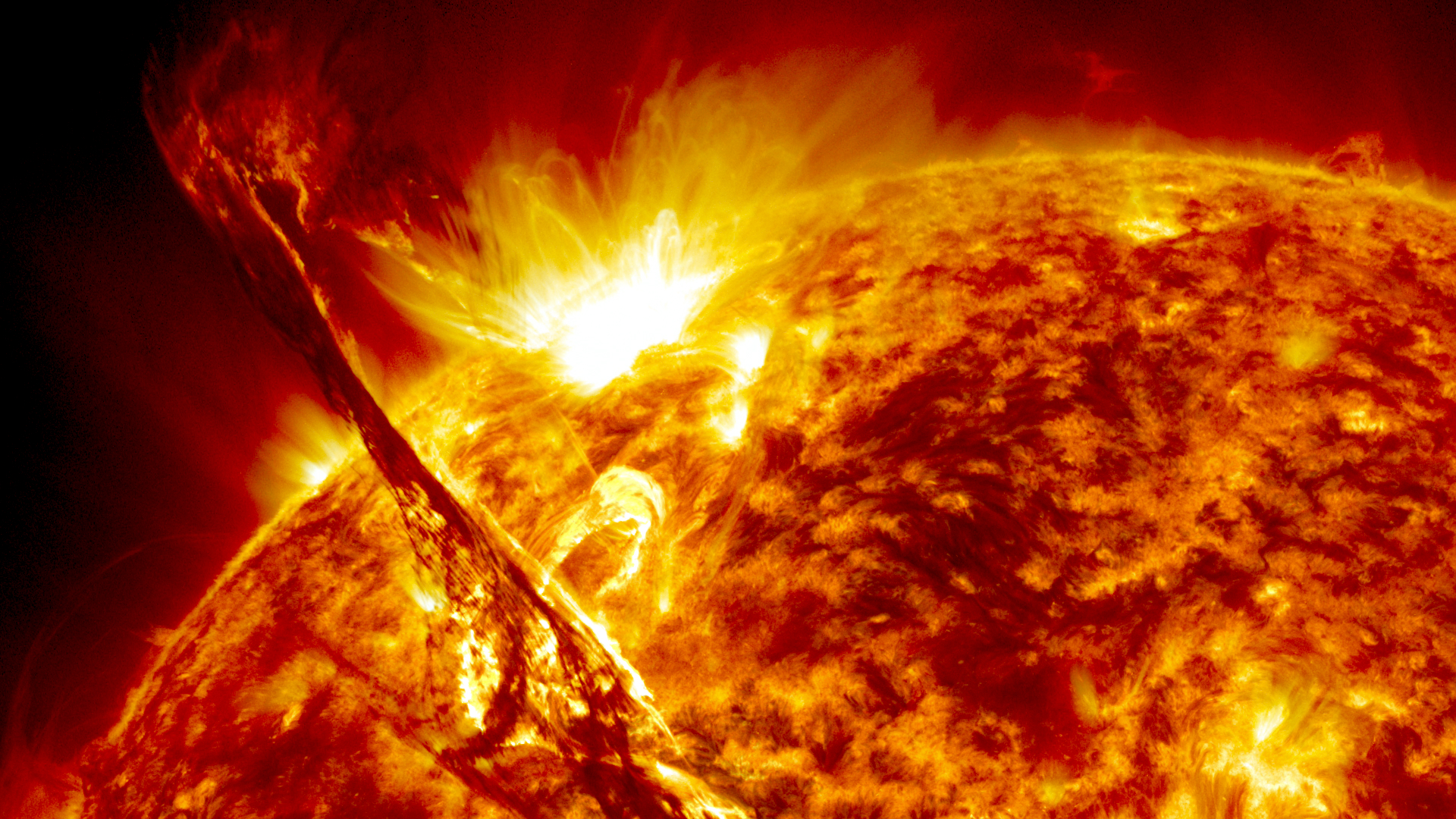}\\
    \footnotesize\textsf{Flare and eruptive prominence leading to a Coronal Mass Ejection as observed in the EUV with the Solar Dynamics Observatory. Credit: SDO/NASA.}
\end{figure}

%----------------------------------------------------------------------
\clearpage
\setstretch{1.0}
\setstretch{0.98}
%\vspace*{-3mm}
\ \\[-12mm]
\begin{tcolorbox}[boxsep=3pt,left=0pt,right=0pt,top=1pt,bottom=1pt,colframe=boxcolor,colback=boxcolor]
\paragraph{Summary.}
% Here we could a provide a short summary 
Societal dependence on space-based services demands major advances in predicting the impacts of eruptive solar events. 
Millimeter-wavelength observations offer uniquely direct access to the time-dependent physical conditions in the atmospheric layers of the Sun where these events originate. 
A facility capable of full-disk, high-cadence, multi-frequency imaging would provide a transformative view of the Sun and its influence on the heliosphere. 
AtLAST is ideally suited to deliver this capability, and to establish a European leadership role in advancing the scientific foundations that will enable reliable, operational space-weather forecasting for the first time.
\end{tcolorbox}

%ALMA already gives unique views of the solar chromosphere in the millimeter regime, probing temperatures and densities directly via thermal emission. The ALMA upgrade would fill some observational gap in the chromosphere (chromospheric gap between photospheric magnetograms and coronal models) and complement ground-based networks like GONG or DKIST. This would enable more accurated observations for providing valuable insight into flare precursor (kernel brightnenings) related to starting reconnection processes on small-scale level. Polarization enables maybe direct measurements of the chromospheric magnetic field? 

\enlargethispage{3mm}
\begin{figure}[b!]
\vspace{-2mm}
\floatbox[{\capbeside\thisfloatsetup{capbesideposition={right,top},capbesidewidth=4.8cm}}]{figure}[\FBwidth]
{\vspace*{1mm}\caption{Space weather events can severely affect modern society because many essential technologies, including satellite-based communication and navigation systems and terrestrial power grids, are highly vulnerable to their impacts. Effective mitigation requires reliable operational space weather forecasting. Achieving this necessitates a deeper understanding of the solar drivers of space weather, their underlying physical mechanisms, and potential early indicators of activity. \\ Image:~ESA/Science~Office.\\
{\small (License: CC BY-SA 3.0 IGO.)}}}
{\includegraphics[width=11.8cm]{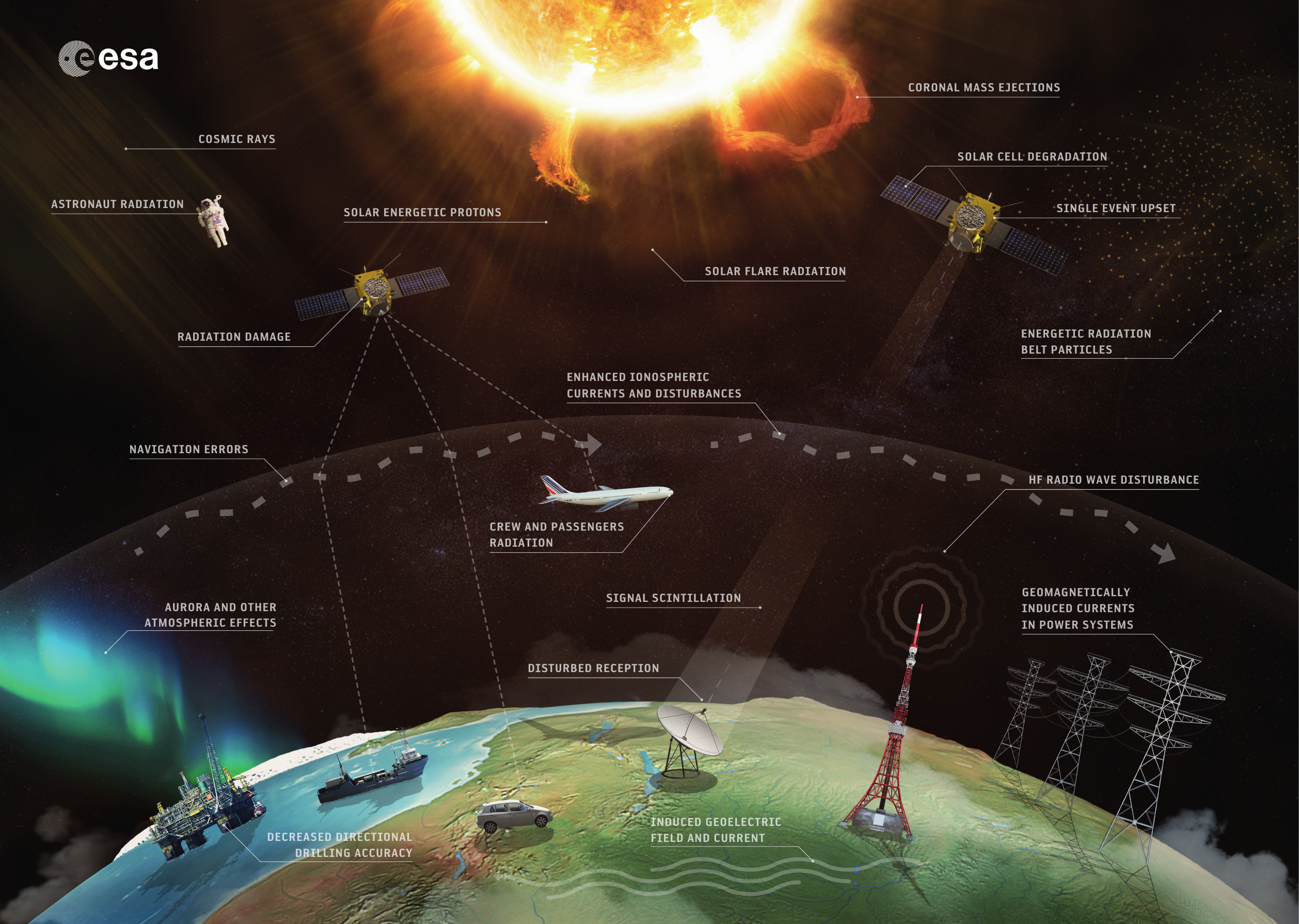}}
\end{figure}

\section{Scientific Context and Open Questions}

\textbf{Space weather} encompasses all solar processes that affect the near-Earth environment and that can significantly influence the operation of modern technological systems  \citep{E-SWAN_2024_SpaceWeatherSpaceClimate}. 
Disruptions to satellite operations, communication and navigation services, power grid stability, and aviation safety pose substantial risks to our society \nocite{2014NHESS..14.2749S,2017SpWea..15..828P,2019SpWea..17.1427B,2024JSpRo..61.1412P}
\citetext{\citenum{2014NHESS..14.2749S}--\citenum{2024JSpRo..61.1412P}}.
Economic assessments estimate that a major space-weather event could incur global costs of tens or even hundreds of billions of euros \citep{2017RiskA..37..206E}. 
\textbf{The urgency of developing reliable forecasting capabilities is therefore undisputed}~\citep{2025EP&S...77..180L}.

Solar flares, eruptive prominences, coronal mass ejections (CMEs), solar energetic particles (SEPs), and the continuously outflowing solar wind are the main drivers of space weather \citep{2021LRSP...18....4T}. 
They originate from a complex interplay of plasma and magnetic fields in the solar atmosphere, spanning from the photosphere through the chromosphere and transition region into the corona. 
Despite major progress over recent decades, we still lack a complete understanding of how energy is accumulated in active regions, how instability and reconnection trigger eruptions, and how rapidly evolving thermal and magnetic structures govern particle acceleration and mass flows into interplanetary space. 
Addressing these knowledge gaps is fundamental to advancing operational space-weather prediction.

\vspace{-3mm}
\paragraph{The Role of Millimeter-Wavelength Solar Observations.} 
The continuum at millimeter (mm) and submillimeter wavelengths provides a unique diagnostic opportunity for measuring the thermal and magnetic structure and dynamics of the chromosphere -- an atmospheric layer of crucial importance for the Sun's activity and thus the sources of space weather.  
Importantly, the mm continuum intensity can be interpreted as a linear measure of the local electron temperature under conditions close to local thermodynamic equilibrium (LTE), in marked contrast to more complex ultraviolet (UV) and extreme ultraviolet (EUV) diagnostic techniques.  
At the same time, the polarisation of the mm continuum provides fundamental information on the line-of-sight magnetic field -- essential information that is otherwise hard to measure. 

The chromosphere, however, is highly structured and dynamic. 
Its magnetic connectivity controls how energy is released in active regions and couples to the overlying corona, enabling or suppressing eruptions. 
Traditional observations in the optical and EUV bands are challenging due to non-LTE effects that hamper the reliable derivation of the required plasma properties. 
Flares, shock waves, and CME onsets all cause strong emission enhancements, offering sensitive diagnostics of particle acceleration and heating mechanisms. 
Observations in the mm domain, therefore, probe the atmospheric layer where space-weather events form and evolve.

So far, the Atacama Large Millimeter/submillimeter Array (ALMA) has pioneered high-resolution mm observations of the Sun. 
It has revealed substantial fine structure in active regions, dynamic chromospheric jets and shocks, and flare-induced heating patterns at high cadence. 
Yet, the current capabilities are limited by small instantaneous fields of view, limited frequency coverage, no simultaneous observing across multiple frequency bands, and the inability to observe the full solar disk at sufficient cadence. 
One could argue that ALMA has revealed the enormous diagnostic potential of the mm continuum for studying the Sun \citep{2016SSRv..200....1W,2018Msngr.171...25B}, and, at the same time, revealed the technological shortcomings that prevent the exploitation of its full scientific potential. \linebreak
\textbf{The potential of mm diagnostics for space-weather science remains largely untapped.}  

\vspace{-3mm}
\paragraph{Tomographic Reconstruction of the Time-Dependent 3D Atmospheric Structure.} 
%-- Unlocking of Diagnostic Potential.} 
As the opacity and thus the effective formation height of the mm continuum depend strongly on wavelength, observing solar emission across the mm wavelength range allows 
mapping the solar atmosphere as a function of height. 
Observing the whole wavelength range strictly simultaneously, which is not currently possible, would enable the tomographic reconstruction of the 3D thermal and magnetic structure of the solar atmosphere as a function of time \citep{2024ORE.....4..140W}. 
This unprecedented tomographic technique would enable continuous quantification of the evolving plasma state in the layers that regulate energy release in active regions, revealing how magnetic energy accumulates, how small changes in topology precede flares and coronal mass ejections, and how impulsive heating modifies the lower atmosphere during eruptions. 
Such data would also help identify when and where solar energetic particles are accelerated and how chromospheric properties influence their release into interplanetary space. 
For the outflowing solar wind, the spatial and temporal variability of its source regions would become trackable with unprecedented diagnostic clarity. 
By providing direct physical information at the interface between the photosphere and corona, this tomographic approach would bridge the gap between sub-surface helioseismic measurements and coronal remote-sensing observations, thereby establishing the missing connection needed for 
accurate modelling of the solar eruptive processes and thus for 
reliable, real-time space-weather forecasting.

\section{Key Science Goals for Space-Weather Research}

Fundamental questions about the solar origins of space weather can only be addressed through dedicated mm-wavelength observations that complement other diagnostics. 
A central challenge is to understand how magnetic reconnection deposits energy into the atmospheric plasma, which then responds through impulsive heating, mass motions, and particle acceleration. 
The development of precursors to eruptions remains elusive because their signatures occur in the dynamically evolving chromosphere, where traditional diagnostics lack the capability to determine the local temperature and magnetic structure with sufficient reliability or cadence. 
During the initiation of coronal mass ejections, the destabilisation of the atmosphere is governed by changes in the thermodynamic structure and magnetic connectivity that mm measurements can capture directly. 
The source regions of solar energetic particles, and the conditions that regulate their acceleration and release, are also rooted in these layers where magnetic field lines open or reconnect. 
Even the origin of the solar wind, a more persistent contributor to space weather, depends on the interplay among temperature gradients, magnetic topology, and small-scale dynamics, which are difficult to capture with optical and EUV instruments alone. 
Addressing these questions is essential for transforming the current descriptive understanding of eruptive solar activity into a genuinely predictive framework.

\section{Technical Requirements}
%A short, less than half a page, description of what technology developments/data handling requirements that may be needed can be included but we are not looking for a detailed descriptions of facilities.

Achieving the above science goals requires a solar observatory that combines full-disk coverage with high temporal and angular resolution in the (sub-)millimeter range, while observing multiple frequencies strictly simultaneously. 
A cadence on the order of one second during eruptive events and several seconds during quiet monitoring is required to track rapid changes that signal imminent eruptions. 
The angular resolution must be sufficiently satisfactory to reveal the spatial signatures of energy release in active regions across the solar disk, and the absolute calibration must be robust enough to interpret observed intensities as local brightness temperatures. 
Crucially, the observational setup must ensure that the solar atmosphere is sampled across a wide range of heights (i.e., at different wavelengths) at a single moment in time, enabling tomographic reconstruction without ambiguities arising from temporal evolution. 
These requirements go beyond the capabilities of current facilities and define the technological advances needed for the next major leap in space-weather diagnostics.

\begin{figure}[t!]
	\vspace*{-2mm}
    \centering
    \includegraphics[width=16.8cm]{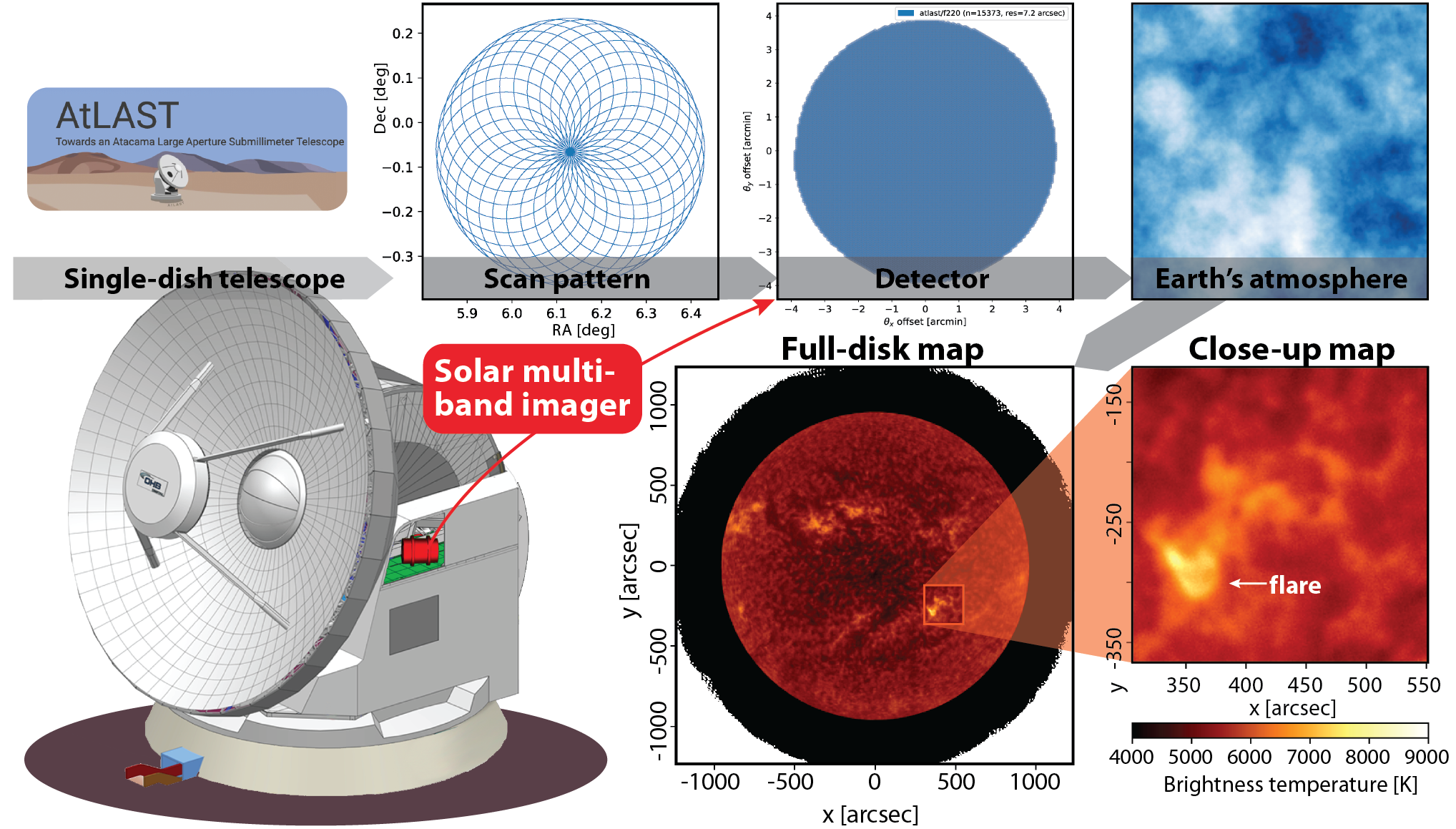}
	%\vspace*{-2mm}    
        \caption{Simulated observations of the Sun with a single frequency (220\,GHz) for a solar multi-band imager installed at AtLAST. A double-circle scan pattern with a multi-pixel detector would enable high-cadence image series of the full disk of the Sun, thereby capturing the origins of space weather, including flares \citep{2025OJAp....8E.129K,2024OJAp....7E.118V}. Such an instrument would provide unprecedented data that enable the 3D tomographic reconstruction of the thermal and magnetic state of the solar atmosphere \citep{2024ORE.....4..140W}. Credits: Adapted from \citep{2025A&A...694A.142M,2025OJAp....8E.129K}. }
    \label{fig:simobs}
\end{figure}

%\clearpage
%\vspace*{5mm}
\textbf{The Atacama Large Aperture Submillimeter Telescope}  \citep[AtLAST,][]{2025A&A...694A.142M} is designed to address exactly these needs. 
Its large single-dish aperture and fast scanning capabilities, combined with advanced multi-band instrumentation, naturally provide the temporal and spatial resolution required for full-disk solar imaging at mm wavelengths in the context of space weather origins (see Fig.~\ref{fig:simobs}). 
The instrument concept for a solar imager foresees the simultaneous use of multiple frequency bands that together span a broad spectral range (at least 30--700\,GHz), enabling a tomographic observations across the entire chromosphere. 
The reliable measurement of the chromospheric magnetic field -- essential for the scientific objective -- requires circular polarisation  or preferably full-Stokes capabilities. 
Monitoring of the chromosphere (and low corona) would thus become possible, including the ability to capture transient events in real time and to remove the pointing challenges imposed by a small field of view. 
Through this capability, AtLAST would deliver the essential measurements that define the thermal and magnetic conditions at the origins of space weather. 

In synergy with other major solar facilities, such as ALMA\footnote{While the spatial resolution of a single-dish telescope like AtLAST will remain below what an interferometric array such as ALMA can achieve, the full-disk, multi-band capabilities make AtLAST and ALMA complementary, with AtLAST an advantage in observing essential phenomena such as flare and CMEs that are hard to predict and point at.}, DKIST, EST, and space missions (ESA's Vigil and successors of SDO and SolarOrbiter), AtLAST would place ESO at the forefront of global efforts to \textbf{understand and forecast solar activity with implications for exoplanet habitability}.  
Its data would ultimately enable the development of reliable, physics-based space-weather prediction tools, which are urgently needed to protect both space- and ground-based technological infrastructure that is increasingly vulnerable to solar disturbances.

\bibliographystyle{phaip}
\begin{multicols}{2}

\end{multicols}

\end{document}